\documentstyle[twoside,fleqn,espcrc2bis,epsf]{article}
\newcommand{\nn}{\nonumber}
\newcommand{\be}{\begin{equation}}
\newcommand{\ee}{\end{equation}}
\newcommand{\bea}{\begin{eqnarray}}
\newcommand{\eea}{\end{eqnarray}}

\def\bfnabla{\mbox{\boldmath $\nabla$}}
\def\bfSigma{\mbox{\boldmath $\Sigma$}}
\def\bfsigma{\mbox{\boldmath $\sigma$}}
\def\bfPi{\mbox{\boldmath $\Pi$}}
\def\lQ{\Lambda_{\rm QCD}}
\def\al{\alpha}
\def\als{\alpha_{\rm s}}
\def\siml{{\ \lower-1.2pt\vbox{\hbox{\rlap{$<$}\lower6pt\vbox{\hbox{$\sim$}}}}\ }}

\newcommand{\Appendix}[1]%
    {%
     \section{#1}%
      }

\begin{document}
\def\siml{{\ \lower-1.2pt\vbox{\hbox{\rlap{$<$}\lower6pt\vbox{\hbox{$\sim$}}}}\ }} 
\def\bfnabla{\mbox{\boldmath $\nabla$}}
\def\bfgamma{\mbox{\boldmath $\gamma$}}
\def\bfSigma{\mbox{\boldmath $\Sigma$}}
\def\bfsigma{\mbox{\boldmath $\sigma$}}
\def\als{\alpha_{\rm s}}
\def\al{\alpha}
\def\lQ{\Lambda_{\rm QCD}}
\def\vs{V^{(0)}_s}
\def\vo{V^{(0)}_o}
\newcommand{\ttbs}{\char'134}
\newcommand{\AmS}{{\protect\the\textfont2 A\kern-.1667em\lower.5ex\hbox{M}\kern-.125emS}}

\title{Poincar\'e invariance constraints on non-relativistic effective field theories}

\author{Antonio Vairo\address{
Institut f\"ur Theoretische Physik, Universit\"at Heidelberg\\ 
Philosophenweg 16, D-69120 Heidelberg, FRG}
\thanks{Alexander von Humboldt fellow}}

\begin{abstract}
We discuss Poincar\'e invariance in the context of non-relativistic effective field theories of QCD. 
We show, in the cases of the HQET and pNRQCD, that the algebra of the generators of the Poincar\'e 
transformations imposes precise constraints on the form of the Lagrangian.
In the case of the HQET they are the relations formerly obtained by reparametrization invariance.
\end{abstract}

\maketitle

\section{INTRODUCTION}
The heavy quark effective field theory (HQET) is the effective field theory of QCD suitable 
for describing heavy-light mesons \cite{HQET}. The heavy-quark four momentum can be 
split into a large and a small component: $p = m v + k$, 
where $m$ is the heavy quark mass, $v$ a unit vector and $k$ a residual momentum 
of the order of $\lQ$. The large and small component fields $\psi_v^\pm$ are defined 
in terms of the heavy-quark field $\Psi$ as 
$\psi_{v}^\pm(x)  = \displaystyle {1 \pm v \!\!\!/ \over 2} e^{im v\cdot x} \Psi(x)$.
The HQET is obtained by integrating out the small component field $\psi_v^-$.
The HQET Lagrangian ${\cal L}$ depends on $v$ either explicitly 
or via the heavy-quark field $\psi_v^+$. 
This dependence is fictitious since $v$ is just a parameter that arbitrarily defines 
the way one splits the heavy-quark momentum into a large and a small part.
Therefore, ${\cal L}$ must be a combination of 
reparametrization invariant operators, i.e. non-reparametrization 
invariant operators must appear in special combinations inside the HQET Lagrangian.
As a consequence, the corresponding Wilson coefficients satisfy 
some exact relations valid at any order in perturbation theory 
\cite{maluke}.

In the following I will be concerned with the relation between reparametrization 
and Poincar\'e invariance. In the first part I will show how to derive the same relations obtained 
from reparametrization invariance, by imposing the Poincar\'e algebra on the generators 
of the Poincar\'e transformations of the HQET, i.e. without introducing the parameter $v$.
The method may be extended to non-relativistic effective field theories (EFTs)
where the relation of the effective degrees of freedom with the original ones 
is less known than in the case of the HQET.
As an application, in the second part I will derive the constraints imposed by Poincar\'e 
invariance on potential NRQCD (pNRQCD).

\section{POINCAR\'E INVARIANCE}
For any Poincar\'e invariant theory the generators $H,{\bf P},{\bf J},{\bf K}$ 
of time translations, space translations, rotations, and Lorentz
transformations satisfy the Poincar\'e algebra:
\begin{eqnarray}
[{\bf P}^i,{\bf P}^j] &=&0, \label{A1}\\ 
{[{\bf P}^i,H]} &=& 0, \label{A2}\\ 
{[{\bf J}^i,{\bf P}^j]} &=& i \epsilon_{ijk}{\bf P}^k , \label{A3}\\  
{[{\bf J}^i,H]} &=& 0, \label{A4}\\  
{[{\bf J}^i,{\bf J}^j]} &=& i \epsilon_{ijk}{\bf J}^k , \label{A5}\\  
{[{\bf P}^i,{\bf K}^j]} &=&-i \delta_{ij} H , \label{A6}\\  
{[H,{\bf K}^i]} &=&-i {\bf P}^i , \label{A7}\\  
{[{\bf J}^i,{\bf K}^j]} &=& i \epsilon_{ijk}{\bf K}^k , \label{A8}\\  
{[{\bf K}^i,{\bf K}^j]} &=&-i \epsilon_{ijk}{\bf J}^k . \label{A9}
\end{eqnarray}
It has been pointed out, as early as in Ref. \cite{Dirac}, that the algebra 
induces non trivial constraints on the form of the Hamiltonian of
non-relativistic systems where Poincar\'e invariance is no longer explicit.
Indeed, the algebra has been used in the past to constrain the form of 
the relativistic corrections to phenomenological potentials \cite{Foldy}. 
A derivation of the constraints induced by Poincar\'e invariance on the form 
of the potential of a quantum-mechanical two-body system can be found in \cite{bgv}.\footnote{
The system studied in \cite{bgv} corresponds to pNRQCD in the non-perturbative regime 
\cite{m1,m2}.}

In a relativistic field theory the fields are representations of the Poincar\'e group and 
the form of the Poincar\'e generators may be derived from the symmetric 
energy-momentum tensor \cite{weinberg}.\footnote{
If $\Theta^{\mu\nu}$ is the symmetric energy-momentum tensor, then 
\begin{eqnarray*}
&& H = \int d^3x  \,\, \Theta^{00}, \qquad 
{\bf P}^i = \int d^3x  \,\, \Theta^{i0}, \\
&& {\bf J}^i = {1\over 2} \int d^3x  \,\, \epsilon_{ijk} \, 
({\bf x}^j \Theta^{k0} - {\bf x}^k \Theta^{j0}), 
\\
&& {\bf K}^i = -t \,{\bf P}^i + \int d^3x \,\, {\bf x}^i \,\Theta^{00}.
\end{eqnarray*}
}
For instance, the  Poincar\'e generators of QCD are given by:
\bea 
H &=& \int d^3x\, \bar{\psi}\,(-i\bfgamma \cdot {\bf D} + m)\,\psi 
\nn\\
&&
+ {\bfPi^{a\, 2}
+ {\bf B}^{a\, 2} \over 2}, \\
{\bf P} &=& \int d^3x\, \psi^\dagger \, (-i{\bf D}) \, \psi 
+ {1\over 2} \, [{\bfPi}^a \times, {\bf B}^a], 
\\
{\bf J} &=& 
\int d^3x\, \psi^\dagger \Bigg({\bf x} \times (-i{\bf D}) + {\bfSigma \over
  2}\Bigg) \psi 
\nn\\
&& 
+ 
{1\over 2} {\bf x} \times [\bfPi^{a}\times, {\bf B}^{a}],\\
{\bf K} &=& -t \, {\bf P} 
\nn\\
&&
+ 
\int d^3x\, \psi^\dagger \, {1\over 2} 
\Bigg\{ {\bf x}, \bar{\psi}\,(-i\bfgamma \cdot {\bf D} + m)\,\psi 
\nn\\
&&
+ {\bfPi^{a\, 2}
+ {\bf B}^{a\, 2} \over 2} \Bigg\},
\label{kqcd}
\eea
where $\bfPi^i_a$ is the canonical variable conjugated to ${\bf A}_{i\,a}$. 
All generators are defined up to a unitary transformation.

In the following I will discuss the realization of the algebra (\ref{A1})-(\ref{A9}) 
in the HQET and pNRQCD, summarizing the findings of \cite{bgv2}.
In a non-relativistic EFT, invariance under Lorentz transformations is  
explicitly broken. The Lorentz-boost generators of the EFT 
can be constructed by matching order by order with the Lorentz-boost generators of QCD 
(\ref{kqcd}). Rotation and space translation are instead preserved symmetries in the EFT, 
therefore the generators are exactly known and are the generators of QCD projected 
on the Hilbert space of the EFT. Since the EFT is equivalent to QCD, i.e. a relativistic 
field theory, the Poincar\'e algebra generators constructed in this way 
must still satisfy the Poincar\'e algebra (\ref{A1})-(\ref{A9}). 
This induces non trivial constraints on the form of the interaction. More specifically, 
it induces some exact relations among the Wilson coefficients of the EFT.

\section{HQET}
\label{secnrqcd}
We quantize the HQET  in the $A^0 = 0$ gauge. 
The pairs of canonical variables are $(\psi, i \psi^\dagger)$,  
$(\chi, i \chi^\dagger)$ and $({\bf A}_{i\, a},  \bfPi^i_a)$.  
The physical states are constrained by the Gau\ss ~law. 

The construction of the generators proceeds in the following way.
The generators $H$, ${\bf P}$ and ${\bf J}$ can be derived from the 
symmetric energy-momentum tensor.
Since translational and rotational invariance 
remain  exact symmetries when going to the effective theory,  
the transformation properties of the new particle fields 
under these symmetries are the same as in the original theory.
The derivation of the Lorentz-boost generators is more problematic, 
since the non-relativistic expansion has destroyed the manifest covariance 
under boosts. A consistent way to construct ${\bf K}$ is to write down the
most general expression, containing all operators consistent with its symmetries 
and to match it to the QCD Lorentz-boost generators.
Accordingly, matching coefficients, typical of ${\bf K}$, will appear.

We obtain:
\bea
h \!\!\!\! &\equiv& \!\!\!\! \psi^\dagger \Biggl( m  - c_1 \, {{\bf D}^2\over 2 m}
- c_2 \, {{\bf D}^4\over 8 m^3}
- c_F\, g {{\bf \bfsigma \cdot B} \over 2 m}
\nn\\
&& \!\!\!\!
- c_D^\prime \, g { \left[{\bf D} \cdot, {\bfPi} \right] \over 8 m^2}
- i c_S \, g { {\bfsigma \cdot \left[{\bf D} \times, \bfPi \right] }
\over 8 m^2} + \dots \Biggr) \psi 
\nn \\ 
&& \!\!\!\!
+ {{\bfPi}^a \cdot {\bfPi}^a  + {\bf B}^a \cdot {\bf B}^a \over 2}
\nn\\
&& \!\!\!\!
- {d_3^\prime\over m^2}\, f_{abc} \, g F^a_{\mu\nu} F^b_{\mu\al} F^c_{\nu\al}
\Bigg|_{\hbox{${\bf E}=\bfPi$}} + \dots ,
\nn \\
\nn \\
H \!\!\!\! &=& \!\!\!\! \int d^3x \, h ,
\label{HNRQCD} \\
{\bf P} \!\!\!\! &=& \!\!\!\!
\int d^3x \, \left( \psi^\dagger \, (-i{\bf D})  \, \psi 
+ {1\over 2}
\, [{\bfPi}^a \times, {\bf B}^a]\right),
\label{PNRQCD} \\
{\bf J} \!\!\!\! &=& \!\!\!\!
\int d^3x \, \left(\psi^\dagger \left( {\bf x} \times (-i {\bf D}) 
+ {\bfsigma \over 2}\right)\psi 
\right. \nn\\
&& \left.
+ {1\over 2} 
\, {\bf x} \times [{\bfPi}^a \times, {\bf B}^a]\right),
\label{JNRQCD} \\
{\bf K}\!\!\!\!  &=& \!\!\!\!
- t \, {\bf P} + \int d^3x \, {\{ {\bf x}, h \} \over 2} 
\nn\\
&& \!\!\!\!
- k^{(1)} 
\int d^3x \, {1\over 2m} \psi^\dagger \, {\bfsigma \over 2} \times 
(-i{\bf D})  \, \psi + \dots. 
\label{KNRQCD}
\eea
The one-loop expressions in the $\overline{{\rm MS}}$ scheme 
for the coefficients $c_F$, $c_D^\prime$, $c_S$ and $d_3^\prime$ 
can be found in \cite{Manohar} (according to the definitions of $ c_D^\prime$ 
and $d_3^\prime$ given in \cite{m2}). The coefficient 
$k^{(1)}$ is a matching coefficient specific of ${\bf K}$.
In principle, $k^{(1)}$ may be calculated at any order 
in perturbation theory by matching (\ref{KNRQCD}) to (\ref{kqcd}).
The tree level value can be also calculated by performing a Foldy--Wouthuysen 
transformation on the Lorentz-boost generators of QCD, in the same way as 
the tree-level matching  coefficients of the HQET Hamiltonian can be derived.
At tree level we have $k^{(1)}=1$.

Let us now consider the constraints induced by the Poincar\'e algebra
(\ref{A1})-(\ref{A9}) on the HQET generators $H$ and ${\bf K}$.
The constraint $[{\bf P}^i,{\bf K}^j] = -i \delta_{ij} H$ 
has been already used in Eq. (\ref{KNRQCD}). 
Indeed, this commutation relation forces ${\bf K}$ to have the form
$\displaystyle \int d^3x$ $\{ {\bf x}, h({\bf x},t) \}/2$ $+$ 
translational-invariant terms.
From $[{\bf K}^i,{\bf K}^j] = -i \epsilon_{ijk}{\bf J}^k$ at ${\cal O}(1/m^0)$ 
it follows that
\be  
k^{(1)}=1.
\label{constrNRQCD1}
\ee
From $[H,{\bf K}^i] = -i {\bf P}^i$ at  ${\cal O}(1/m^0)$ it follows that 
\be  
c_1=1,
\label{constrNRQCD2}
\ee
and at  ${\cal O}(1/m)$ 
\be
2c_F -c_S -1 =0.
\label{constrNRQCD3}
\ee
Finally from $[H,{\bf K}^i] = -i {\bf P}^i$ at  ${\cal O}(\bfnabla^2 \,
\bfnabla^i /m^2)$ we obtain 
\be  
c_2=1.
\label{constrNRQCD4}
\ee
All other commutation relations are satisfied at the order we are working. 
The constraints (\ref{constrNRQCD2}), (\ref{constrNRQCD3}) and (\ref{constrNRQCD4})
were first derived in the framework of reparametrization invariance 
in \cite{maluke,Manohar}.

\section{pNRQCD}
\label{secpnrqcd}
The pNRQCD Lagrangian for a heavy quark-antiquark system is obtained from the 
NRQCD Lagrangian \cite{NRQCD} by integrating out the soft degrees of freedom associated with the scale
of the relative momentum of the two heavy quarks 
in the bound state \cite{pnrqcd}. The name pNRQCD has been used 
in the literature to identify effective field theories with different degrees 
of freedom. Here we call pNRQCD the effective field theory that can be
obtained from NRQCD by perturbative matching and contains, as degrees of
freedom, the quark-antiquark field (that can be split into 
a colour singlet ${\rm S} = { S 1\!\!{\rm l}_c / 
\sqrt{N_c}}$ and a colour  octet ${\rm O} = O^a { {\rm T}^a / \sqrt{T_F}}$ component)
and (ultrasoft) gluons. The fields $S$ and $O^a$ are functions of 
$({\bf X},t)$ and  ${\bf x}$, where ${\bf X}$ 
is the centre-of-mass coordinate and ${\bf x}$ the
relative coordinate. The coordinate ${\bf x}$ plays the role 
of a continuous parameter, which specifies different fields.
All the gauge fields have been multipole expanded around the centre-of-mass. 
Therefore, the terms in the pNRQCD Lagrangian are organized in powers of $1/m$ and $x$. 

The canonical variables and their conjugates are 
$(S, i S^\dagger)$, $(O_a, i O^\dagger_a)$, and $({\bf A}_{i\,a},\bfPi^i_a)$.
The physical states are constrained to satisfy the Gau\ss ~law.

As in the case of the HQET, since translational and rotational invariance are 
exact symmetries of the effective theory, the generators $H$, ${\bf P}$ 
and ${\bf J}$ can be derived from the symmetric energy-momentum tensor.
The pNRQCD Lorentz-boost generators ${\bf K}$ can be derived by 
writing down the most general expressions containing all operators consistent 
with their symmetries and by matching them to the NRQCD Lorentz-boost generators, which, 
at the order we are working here, are the simple extension of Eq. (\ref{KNRQCD}) 
to the case of one particle and one antiparticle.

We obtain:
\bea
h &\equiv&  {{\bfPi}^{a\,2} + {\bf B}^{a\,2} \over 2} 
+ \int d^3x \, {\rm Tr} \,\Biggl\{ 
  {\rm S}^\dagger (2m + h_S) {\rm S}  
\nn\\
&&
+ {\rm O}^\dagger (2m + h_O) {\rm O}  
\nn\\
&&
+ \left[ ({\rm S}^\dagger h_{SO} {\rm O} + {\rm H.C.}) + {\rm C.C.}  \right] 
\nn\\
&&
+ \left[ {\rm O}^\dagger h_{OO} {\rm O} + {\rm C.C.}  \right] 
\nn\\
&&
+ \left[ {\rm O}^\dagger h_{OO}^A {\rm O}  h_{OO}^B + {\rm C.C.}  \right] 
\Biggr\},
\nn\\
  H &=& \int d^3X \, h,
\label{HpNRQCD}
\\ 
{\bf P} &=& \int d^3X \int d^3x \, {\rm Tr} \,\Biggl\{
{\rm S}^\dagger ({-i \bfnabla}_X) {\rm S} 
\nn\\
&&
+ {\rm O}^\dagger (-i{\bf D}_X) {\rm O} \Biggr\}
\nn\\
&&
+ {1\over 2} \int d^3X \, [{\bfPi}^{a}\times, {\bf B}^{a}],
\label{PpNRQCD}
\\
{\bf J} &=& \int d^3X \int d^3x \, {\rm Tr} \,\Biggl\{
{\rm S}^\dagger \Bigg({\bf X} \times ({-i \bfnabla}_X) 
\nn\\
&&
+ {\bf x} \times (-i {\bfnabla}_x) 
+ {\bfsigma^{(1)}+\bfsigma^{(2)} \over 2}  \Bigg){\rm S} 
\nn\\
&&
+ {\rm O}^\dagger \Bigg({\bf X} \times (-i {\bf D}_X) 
\nn\\
&&
+ {\bf x} \times (-i {\bfnabla}_x) 
+ {\bfsigma^{(1)}+\bfsigma^{(2)} \over 2}  \Bigg) {\rm O} \Biggr\}
\nn\\
&&
+ {1\over 2} \int d^3X \, {\bf X} \times [{\bfPi}^{a}\times, {\bf B}^{a}],
\label{JpNRQCD}
\\
{\bf K} &=& -t \, {\bf P} + \int d^3X \, {1\over 2}\, \{ {\bf X}, h \} 
\nn\\
&&
+  \int d^3X \int d^3x \, {\rm Tr} \,\Bigl\{ 
  \left[ {\rm S}^\dagger {\bf k}_{SS} {\rm S} + {\rm C.C.} \right] 
\nn\\
&&
+ \left[ ({\rm S}^\dagger {\bf k}_{SO} {\rm O} + {\rm H.C.}) + {\rm C.C.} \right] 
\nn\\
&&
+ \left[ {\rm O}^\dagger {\bf k}_{OO} {\rm O} + {\rm C.C.} \right] \Bigr\},
\label{KpNRQCD}
\eea
where  C.C. stands for charge conjugation and H.C. for Hermitian conjugation.
Explicit expressions for $h_{S}$, $h_{O}$, $h_{SO}$, $h_{OO}$ and $h_{OO}^{A,B}$ 
at order $x^2/m^0$, $x^0/m$, $(x/m) \, {\bfnabla}_X$ and  $(x^0/m^2) \, {\bfnabla}_X$ 
and for ${\bf k}_{SS}$, ${\bf k}_{SO}$ and ${\bf k}_{OO}$ 
at order $x^2/m^0$, $x^0/m$ and  $(x/m) \, {\bfnabla}_X$ can be found in \cite{bgv2}. 

Let us consider the constraints induced by the Poincar\'e algebra
(\ref{A1})-(\ref{A9}) on the pNRQCD generators $H$ and ${\bf K}$.
The constraint $[{\bf P}^i,{\bf K}^j] = -i \delta_{ij}
H$ has been already used in writing Eq. (\ref{KpNRQCD}). 
Indeed, it forces ${\bf K}$ to have the form 
$\displaystyle \int d^3X \, \{ {\bf X}, h({\bf X},t) \}/2 \,+\,$ 
translational-invariant terms.
From the constraints that we get on ${\bf K}$ from the other commutation 
relations and using the freedom that we have to 
redefine the Poincar\'e generators via a unitary transformation 
we obtain at the order we are working:
\bea
{\bf k}_{SS} &=& 
- {1\over 4m} \left(\bfsigma^{(1)} \times (-i\bfnabla_x) \right)
\nn\\
&&
- {1\over 8m} \{ {\bf x}, {\bfnabla}_X \cdot \bfnabla_x\}, 
\\
{\bf k}_{OO} &=& 
- {1\over 4m} \left(\bfsigma^{(1)} \times (-i\bfnabla_x) \right)
\nn\\
&&
- {1\over 8m} \{ {\bf x}, {\bf D}_X \cdot \bfnabla_x\}
\nn\\
&&
- {1\over 8} k_{OO a}^{(0,2)}(x) \, {\bf x} \,({\bf x}\cdot g \bfPi) 
\nn\\
&&
- {1\over 8} k_{OO b}^{(0,2)}(x) \, {\bf x}^2 \, g \bfPi, 
\\
{\bf k}_{SO} &=& 0,
\eea
where $k_{OO a}^{(0,2)}$  and $k_{OO b}^{(0,2)}$ are some 
matching coefficients specific of ${\bf K}$.
At tree level we have  $k_{OO a}^{(0,2)}=1$ and  $k_{OO b}^{(0,2)}=0$. 
The Poincar\'e algebra also constrains the form of the 
pNRQCD Lagrangian. In the following, I will discuss the different type 
of constraints. 

{\bf (A) Kinetic energy.}  The centre-of-mass kinetic energy is fixed 
to be equal to $ - {\bfnabla}_X^2/4m$. We note that Poincar\'e invariance by itself 
does not constrain the coefficient of the kinetic energy of the quarks in the centre-of-mass frame.

{\bf (B) Potentials of order $1/m^2$.} If we call $V^{(0)}$ the static potential, 
$V_{{\bf p}^2}$ the $1/m^2$ momentum square dependent potential, $V_{{\bf L}^2}$ 
the $1/m^2$ angular momentum square potential and $V_{LS}$ the $1/m^2$ spin-orbit potential 
either in the singlet ($S$) or in the octet ($O$) sector we obtain:
\bea
&& \hspace{-5mm} 
{V_{LS\, Sa} \over V_S^{(0)\prime}} = {V_{LS\, Oa} \over V_O^{(0)\prime}}
= -\frac{1}{2x},
\label{ex0}
\\
&& \hspace{-5mm} 
V_{{\bf L}^2\, Sa} + {x \, V_S^{(0)\prime} \over 2}
=
V_{{\bf L}^2\, Oa} + {x \, V_O^{(0)\prime} \over 2}
= 0,
\label{ex0bis}
\\
&& \hspace{-5mm} 
V_{{\bf p}^2\, Sa} + V_{{\bf L}^2\, Sa} + {V_S^{(0)} \over 2}
=
\nn\\
&& \hspace{10mm} 
V_{{\bf p}^2\, Oa} + V_{{\bf L}^2\, Sa} + {V_O^{(0)} \over 2}
= 0,
\label{ex0tris}
\eea
where the label $a$ is kept for consistency with the notation of \cite{bgv2}.
Eq. (\ref{ex0}) in the singlet sector is the relation between the spin-orbit potentials 
and the static potential first derived in \cite{poG}. Eqs. (\ref{ex0bis})-(\ref{ex0tris}) 
in the singlet sector are the relations between the
momentum-dependent potentials first derived in \cite{poBBP}.
They were also  obtained in \cite{bgv}. A lattice check of these 
relations has been done in \cite{bali}.
The extension to the octet sector has been derived in \cite{bgv2}.

{\bf (C) Singlet and octet couplings to gluons.} In the sector of the pNRQCD 
Lagrangian containing the couplings of the heavy-quarkonium fields to the 
gluons we may distinguish two set of relations induced by the Poincar\'e invariance. 
The first set constrains the chromoelectric field to enter the Lagrangian just in the combination
\bea 
&&{\bf x}\cdot \left(g{\bf E} + 
{1\over 2} \left\{ {-i{\bf D}_X\over 2m} \times, g {\bf B}\right\} \right),
\label{exx0}
\eea 
i.e. like in the Lorentz force.
The second set contains relations that involve combinations of matching coefficients 
appearing at different orders in the expansion in $1/m$ and $x$:
\bea 
&& \hspace{-5mm} {2\, c_F V_{SOb}^{(1,0)} -c_s V_{SOa}^{(2,0)} \over  V_{SO}^{(0,1)}}
=  2 
{c_F V_{OOb}^{(1,0)} + V_{O\otimes Ob}^{(1,0)}\over
V_{OO}^{(0,1)}}
\nn\\
&& \hspace{5mm}
- {c_s  V_{OOa}^{(2,0)} + V_{O\otimes Oa}^{(2,0)} \over
V_{OO}^{(0,1)}}
= 1,
\label{ex1}
\\
&& \hspace{-5mm} 2\, V_{SOc}^{(1,0)} - V_{SOb'}^{(2,0)}  = 
2\, \left(V_{OOc}^{(1,0)} + V_{O\otimes Oc}^{(1,0)} \right)  
\nn\\
&& \hspace{5mm}
- \left(V_{OOb'}^{(2,0)} + V_{O\otimes Ob'}^{(2,0)} \right)
= 0,
\label{ex2}
\\
&& \hspace{-5mm} {-V_{SOb''}^{(2,0)} \over 
x \, V_{SO}^{(0,1)\prime} }
= 
{- V_{OOb''}^{(2,0)} - V_{O\otimes Ob''}^{(2,0)} 
\over x \, V_{OO}^{(0,1)\prime}}
= 1,
\label{ex3}
\\
&& \hspace{-5mm} 
V_{SOb'''}^{(2,0)}
= V_{OOb'''}^{(2,0)} + V_{O\otimes Ob'''}^{(2,0)} 
= 0,
\label{ex4}
\\
&& \hspace{-5mm} 
V_{OOa}^{(1,0)} =1 + {V_{OOc''}^{(2,0)} - V_{OOc'}^{(2,0)} \over 2},
\label{ex5}
\eea 
where $V^{(i,j)}$ are the matching coefficients that appear at order $x^j/m^i$, 
the letters $a$, $b$, ... label different operators explicitly listed in \cite{bgv2}
and the specifications $SO$ and $OO$ refer to couplings with gluons in the singlet-octet 
and octet-octet sector respectively. Eq. (\ref{ex1}) involves  combinations of 
matching coefficients inherited from HQET/NRQCD. Somehow this
relation reflects at the level of the pNRQCD matching coefficients the relation
(\ref{constrNRQCD3}) among the HQET matching coefficients.
Eq. (\ref{ex3}) involves derivatives of $V_{SO}^{(0,1)}$ and $V_{OO}^{(0,1)}$.
Eq. (\ref{ex5}) is typical for the non-Abelian structure of QCD.

\section{OUTLOOK}
In this contribution I have addressed two questions:
what is the relation between reparametrization invariance and 
Poincar\'e invariance in the HQET; may Poincar\'e invariance be used to constrain 
the form of other non-relativistic effective field theories.

For what concerns the first question, 
in Sec. \ref{secnrqcd} we have shown that by imposing 
the  Poincar\'e algebra on the generators of the Poincar\'e transformations
of the HQET we obtain the relations, Eqs. (\ref{constrNRQCD1})-(\ref{constrNRQCD4}), 
derived formerly from reparametrization invariance.  
This shows that reparametrization invariance is, indeed, 
one way in which the Poincar\'e invariance of QCD manifests itself in the HQET 
(see also \cite{eiras}).

The second question may be of particular interest when very little is known about 
the relation between the effective degrees of freedom and the original ones. 
This is, for instance, the case when dealing with EFTs in the non-perturbative domain.
In Sec. \ref{secpnrqcd} we have outlined the calculation of the constraints induced 
by the Poincar\'e invariance on the form of the pNRQCD Lagrangian.
Some of the obtained relations in the singlet potential sector of the 
pNRQCD Lagrangian were already derived in the literature by explicitly 
boosting their expression in terms of Wilson loop operators. 
The same relations also apply to the octet sector. New  relations involving the couplings 
of the singlet and octet fields with the gluons were 
derived in \cite{bgv2} and are displayed in Eqs. (\ref{exx0})-(\ref{ex5}).

Another, quite natural, non-relativistic EFT where to apply the above method 
would be NRQCD. 
Up to order $1/m$ the NRQCD Lagrangian in the particle and antiparticle 
sector coincide with that one of the HQET and, therefore, the relations 
induced by Poincar\'e invariance are the same. Starting from the order $1/m^2$, 
the NRQCD Lagrangian contains four-fermion operators, which are responsible 
for decay and production processes. Poincar\'e invariance will constrain 
the form of these operators. In general, the presented approach, 
may be suited to derive exact relations among the matching coefficients 
of all effective field theories, where the manifest covariance under boosts has been 
destroyed by an expansion in some small momenta, like the soft-collinear effective theory \cite{repscet}.
\vspace{1mm}\\

{\bf Acknowledgments}

I thank Nora Brambilla and Dieter Gromes for collaboration on the work presented here 
and the Jefferson Lab theory group and Jos\'e Goity for hospitality during the 
writing up.

\end{document}